# Enhanced critical temperatures in the strongly overdoped iron-based superconductors $A$Fe$_2$As$_2$ ($A$ = K, Cs, Rb) observed by point contacts

Yu. G. Naidyuk[1], O. E. Kvitnitskaya[1], D. V. Efremov[2], S.-L. Drechsler[3]

[1]*B. Verkin Institute for Low Temperature Physics and Engineering, NAS of Ukraine, 61103 Kharkiv, Ukraine*
[2]*Institute for Solid State Research, IFW Dresden, D-01171 Dresden, Germany*
[3]*Institute for Theoretical Solid State Physics, IFW Dresden, D-01171 Dresden, Germany*

**Abstract**

A remarkable several times increase (up to 10 K) of the superconducting critical temperature $T_c$ has been observed in point contacts created on the base of single crystals $A$Fe$_2$As$_2$ ($A$ = K, Cs, Rb). Possible reasons for such a $T_c$ increase in point contacts are briefly discussed on a qualitative level. Among them, it is most likely attributed to interfacial carrier doping and/or uniaxial non-homogeneous pressure arising when the contact is created.

**Introduction**

The stoichiometric $A$Fe$_2$As$_2$ ($A$ = K, Cs, Rb) compounds belong to the extremely-hole-doped systems within the rich family of the iron-based superconductors. These compounds differ from their optimally doped counterparts by the presence of strong electron correlations as follows from the reported effective mass enhancement and large Sommerfeld coefficients, see e.g. [1,2,3]. The three compounds KFe$_2$As$_2$, RbFe$_2$As$_2$, and CsFe$_2$As$_2$, with superconducting transition temperatures $T_c$ about 3.8K [4], 2.6K [3] and 1.8K [1] respectively, may have a common SC gap structure, where according to LDA calculations their low-$T_c$ values still cannot be explained by the electron-phonon interaction [4]. The Fermi surface topology remains almost unchanged while being doped by different alkali-metal ions ($A$ = K, Cs, Rb) [5] in spite of some changes in the anisotropy of the Fermi velocities (plasma frequencies) according to DFT calculations [6,7]. The pairing symmetry of these materials may be different from that proposed $s^{\pm}$-scenario for the optimally doped iron-based superconductors [8,9] since the title compounds have only hole-pockets at the Fermi energy. There is a general consensus about the presence of a line-node at least at one of the four hole-Fermi surface sheets. In case of a single one it would correspond to an accidental one (extended $s^{\pm}$-wave) and in case of a global one to a $d$-wave superconducting order parameter. Thus, the pairing mechanism for the highly hole-doped composition remains still under debate [10, 11]. This motivates us to investigate deeper the superconducting state of these systems by different experimental tools and under external impact.

The title materials demonstrate also an intriguing behavior of $T_c$ under external pressure. Tafti *et al.* [12] reported a sudden reversal in the pressure dependence of $T_c$ in these materials at a critical pressure $P_c$ = 1.1 GPa connecting it with a change in the structure of the superconducting gap from *d*- to $s^{\pm}$-wave state, while no any change in the Fermi surface occurs. Instead of this, Nakajima *et al.* [13] reported a high-pressure study of transport and structural properties of $KFe_2As_2$ up to 33 GPa, where in the low-pressure regime, they found that the $T_c$ exhibits a maximum onset value of 7 K near 2 GPa. In addition, the $T_c$ is diminished upon applying higher pressures until a sudden appearance of superconducting phase above 13 GPa, which coincides with a first-order structural phase transition into a collapsed tetragonal phase. They suggested that the high-temperature superconducting phase in $KFe_2As_2$ is substantially enhanced by the presence of nested electron and hole pockets upon Fermi surface reconstruction. Contrary, Wang *et al.* [14] observed that $T_c$ of the tetragonal $KFe_2As_2$ was suppressed gradually and disappears completely at ~11 GPa while no superconductivity appears in the collapsed tetragonal phase of $KFe_2As_2$. The discrepancy between this and the previous reports they attribute to the difference in the pressure homogeneity.

On the other side, Yang *et al.* [15] using STM reported about a characteristic temperature of 13K in $CsFe_2As_2$ at ambient pressure, which is up to four to six times larger than the bulk $T_c$. Yang *et al.* connected this with the possible existence of superconducting fluctuations or a pseudogap state. All above mentioned studies indicate that the physics of the superconducting state in the title compounds is very rich and complex and far to be completely understood.

Point contact (PC) spectroscopy has been widely used last years for studying emergent materials. In addition, the PC itself plays a role not only as a probe or physical instrument, but also novel phenomena arise, because of the small size of the volume studied, the high current density and electric field achieved in the PC [16, 17] along with an important role of the interface. A recent example is, e.g., the observation of enhanced superconductivity in PCs with topological Dirac [18] and Weyl [19] semimetals, where the hunt for topological superconductivity continues. In this case, PC spectroscopy can provide a deeper insight of the exotic and emergent properties of the materials studied.

In the presented paper, we investigate three compounds $AFe_2As_2$ ($A$ = K, Cs, Rb), focusing on the observed several times increase of $T_c$ in PCs based on these iron-based low temperature superconductors. Our observation of strong $T_c$ increase in PCs at the interface with normal metal may give additional details to gain a better insight into the superconducting state and its nature in the title materials.

**Method**

Single crystals of $A$Fe$_2$As$_2$ ($A$ = K, Cs, Rb) were grown using the self-flux method as described in Refs. [3,4].

The PCs were established at helium temperature using home-made insert by touching of thin Cu wire to a cleaved surface of $A$Fe$_2$As$_2$ ($A$ = K, Cs, Rb) flake or contacting by Cu wire edge of plate-like samples. Likewise, so called 'soft' PCs were prepared by putting a tiny drop of silver paint on the freshly cleaved surfaces of the samples at room temperature.

We measured the $dV/dI$ derivative of the current–voltage ($I$–$V$) characteristics of PCs in the temperature range between 2 and 10 K. The $dV/dI(V)$ was recorded by sweeping the dc current $I$ on which a small ac current $i$ was superimposed.

**Results and discussion**

Figure 1 shows $dV/dI(V)$ curves for some selected PCs based on $A$Fe$_2$As$_2$ ($A$ = K, Cs, Rb) compounds measured at different temperatures. We observed sharp zero-bias minimum in $dV/dI$ up to 7 K for some PCs on their basis, which is characteristic for superconducting state, although the investigated compounds have the $T_c$ below 4 K (see Table). The magnitude of the minimum gradually decreases with increase of temperature and it vanishes at a temperature about 7 K (Fig. 1), which is in line with their superconducting origin. Additionally, spikes in $dV/dI$, as seen in Fig.1(c) (less pronounced in Fig.1(a,b)), is hallmark of suppression of superconductivity in PC core by high current density, which can be accompanied also by heating [16,17].

The increase of $T_c$ in PCs could be at first glance connected with the effect of pressure at the PC formation by touching the sample with a Cu wire. However, the study of these systems under the hydrostatic pressure up to 2.5 GPa [12] demonstrates in general a *decrease* of the $T_c$ value as mentioned in the Introduction. Although Nakajima *et al.* [13] found that $T_c$ has a maximum onset value of 7 K near 2 GPa, we assume that metallic Cu wire used as a conterelectrode cannot produce a pressure larger than the yield strength of Cu. The latter reaches only about 0.07 GPa [20] and cannot be much larger at liquid helium temperature. Moreover, presented in the Table "soft" PC with CsFe$_2$As$_2$ (in this case pressure is hardly expected) also demonstrates about two times larger $T_c$. So, the pressure origin of the $T_c$ growth is questionable. Albeit, taking into account a scatter of $T_c$ in our PCs (see Table) and the broad distribution of $T_c$ found in the literature for the title compounds under pressure $P$ and even qualitatively different $T_c$ versus $P$ dependencies, it is not excluded that uniaxial non-homogeneous pressure under the PC tip might induce or enhance superconductivity in PC core at interface.

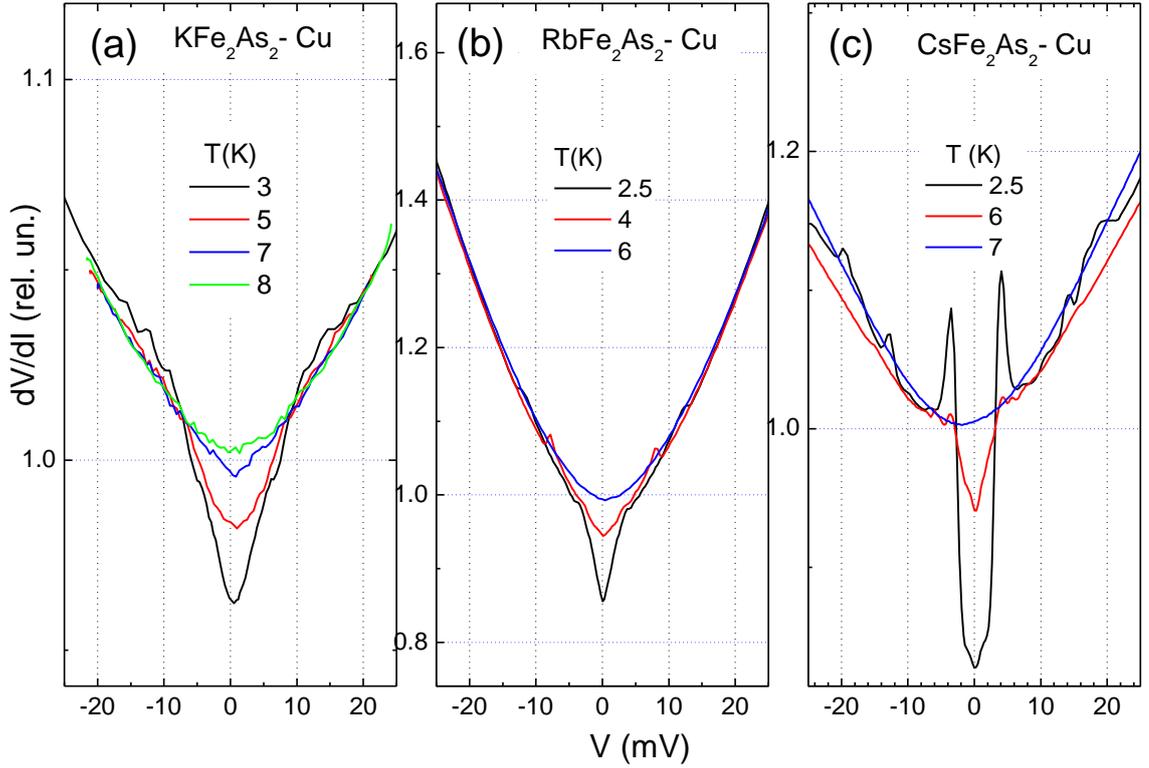

**Figure 1**. Behavior of the differential resistance *dV/dI* of PCs on the base of the $KFe_2As_2$ (a), $RbFe_2As_2$ (b) and $CsFe_2As_2$ (c) at different temperatures. PC resistance is $R_{PC}=25\Omega$ (a), $R_{PC}=4.6\Omega$ (b) and $R_{PC}=5.6\Omega$ (c).

Figure 2 shows more detailed behavior of *dV/dI(V)* curves versus temperature of PC from Fig.1(a). Here, the inset demonstrates the temperature dependence of the integral intensity of zero-bias minimum. Extrapolation of the intensity to zero results in characteristic temperature between 9 and 10 K. This is close to the temperature 9 K at which the zero-bias minimum in STS spectra of $CsFe_2As_2$ disappears as reported by Yang *et al.* [15]. The latter has been designated by Yang *et al.* as pseudogap-related features. From this point of view the investigated compounds exhibit some similarity with strongly correlated high-$T_c$ superconductors, where the pseudo-gap region expands to temperatures several times lager that the bulk $T_c$ [21]. So enhanced superconductivity in the title compounds with pseudo-gap features or enlarged superconducting fluctuation region may indicate its unconventional character and points in our opinion to *d*-wave as a signature for strong correlations. If one excludes a pressure effect, the direct PC might lead to an electron injection of the surface region (interface) retaining this way a less strongly hole-doped region and enhanced $T_c$-values as observed for $Ba_{0.1}K_{0.9}Fe_2As_2$ in, for instance, [22]. In this case an upshift of the characteristic bosonic feature near 20 meV observed in PC spectra in our previous paper [4] might be predicted, which has therefore still not resolved for $CsFe_2As_2$ and $RbFe_2As_2$ samples so far.

Alternatively an enhanced damping caused by the changed anisotropic electron structure and/or additional scattering due to induced inhomogeneities (see below for the evidence of enhanced resistivity) might essentially *broaden* this still not well understood feature hindering the experimental resolution (detection) of the remnant. A microscopic straightforward calculation for the relevant loss function –Im $1/\varepsilon(\omega,q)$ and the corresponding PC counterpart to discuss the possibilities for a peculiar charge-charge response (being of considerable interest for the quasi-2D superconductivity in iron pnictides in general) is left for a special theoretical study in near future. Finally, the presence of the mentioned above boson [4] might be in principle also detrimental for superconductivity by competing with magnetic fluctuations as the main reason for unconventional nodal intraband superconductivity coexisting with unconventional interband contributions, if the latter support a competing $s^{++}$ interband mechanism. This way it might lower the resulting interband coupling. In this case, the mentioned above damping possible could be helpful to rise $T_c$ especially in a weak coupling regime.

At this point we note, that the large increase of *dV/dI* with bias (see $\Delta R$ in the Table) is characteristic for the nonspectral (thermal) regime [23]. This points to enhanced scattering of electrons in PC core and to shortening of their inelastic mean free path. The latter is favored by the reduction of the elastic mean free path, which is governed by lattice defects/impurities in PC core. Thereby, disturbed/disordered lattice at interface in PC may be the cause for the enhanced $T_c$. However, the vast majority of PCs with similar increase of *dV/dI* do not show enhanced $T_c$. Therefore, distorted lattice in PC by itself cannot explain such $T_c$ increase.

Anyhow, in this context we remind the reader that the interplay of disorder and unconventional nodal superconductivity is by no means trivial (see e.g. [24] and Refs. therein) and requires additional theoretical studies for the present case. In this context, the recent observation of a resonance mode in $KFe_2As_2$ strongly supports the unconventional nature of superconductivity [25] although several details of the interpretation remain in our opinion unclear as explained in detail elsewhere.

Interestingly that we have observed similar up to two times increase of $T_c$ in PCs with another iron-based superconductor FeSe [26]. Due to the fact that $T_c$ of a one-unit-cell FeSe film (single layer) is dramatically enhanced [27], it is challenge to test superconductivity in single/multi layers of $AFe_2As_2$ ($A$ = K, Cs, Rb) as well. For the single layer FeSe any details of the expected 2D peculiar charge-charge response have not been considered theoretically so far and therefore any full quantitative description of enhanced superconductivity scenario due to strengthened electron-phonon coupling alone, as widely discussed in the literature, from the coupling to high frequency oxygen phonons derived from the interface with the substrate is not yet convincing in the presence of enhanced/changed Coulomb repulsion. In particular, the use of a weak *repulsive* Coulomb pseudopetential $\mu^* \sim 0.1$ as in standard 3D wide band superconductors seems to be *not* justified here since it rests on the presence of a large plasma frequency (as compared to the high oxygen related phonon frequency of the substrate ~ 60-80 meV) missing here. To conclude, the mechanism of the observed similar $T_c$-enhancement in four bulk superconducting systems remains a challenging puzzle. We speculate that strain and pressure sensitive surface topological superconductivity as in Weyl- semimetal MoTe$_2$ [19] might play a considerable role in all these systems.

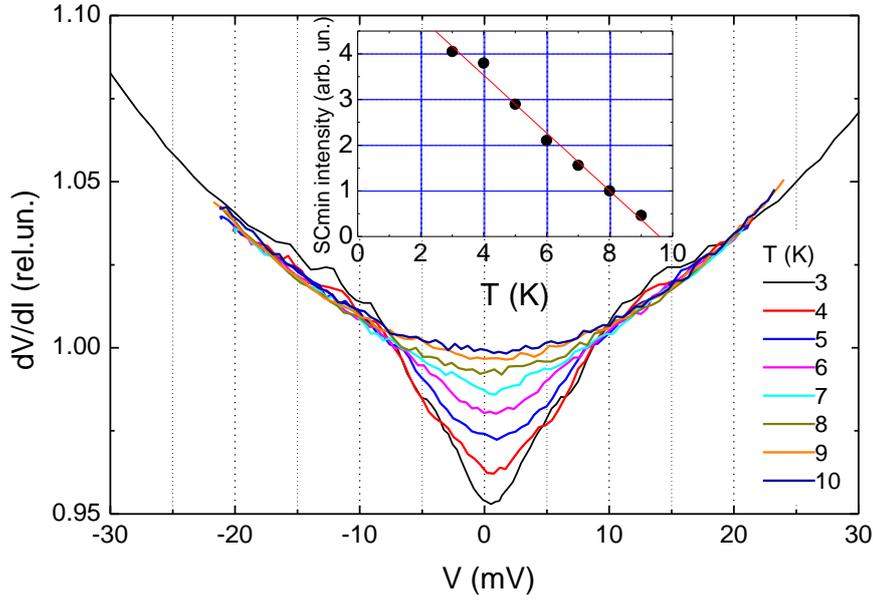

**Figure 2.** Evolution of the differential resistance *dV/dI* of KFe$_2$As$_2$ PC with R=25 Ω versus temperature. Inset: Temperature dependence of the integral intensity of the zero-bias SC minimum for the PC from the main panel (symbols) with its linear approximation (line). The **i**ntegral intensity was calculated after subtracting from each curve its *dV/dI* at 10K.

In conclusion, we observed strongly enhanced critical temperature $T_c$ in PCs created on the base of iron-based superconductors $A$Fe$_2$As$_2$ ($A$ = K, Cs, Rb). Such increase of $T_c$ in PCs is most likely due to potential interfacial carrier doping along with uniaxial non-uniform pressure produced by the PCs formation. Anyway, observed several times increase of $T_c$ in these compounds raises a question about their unconventional superconducting state.

**Acknowledgments**

The authors acknowledge funding by the Volkswagen Foundation. We are grateful to S. Aswartham for samples supply. Yu.G.N. and O.E.K. are thankful for support by the National Academy of Sciences of Ukraine under project Φ4-19 and would like to thank the IFW-Dresden for hospitality.

**Table 1.** Critical temperature of zero-bias minimum suppression $T_c^{PC}$ for several PCs in comparison with the bulk $T_c^{bulk}$. ΔR is relative increase of *dV/dI* at 100 mV.

| Sample | Counter-electrode | $R_{PC}$ (Ω) | $T_c^{PC}$ (K) | $T_c^{bulk}$ (K) | ΔR (%) |
|---|---|---|---|---|---|
| KFe$_2$As$_2$ | Cu | 25 | ~8 | 3.8 | ≈50 |
| | | 2.6 | ~5 | | ≈70 |
| RbFe$_2$As$_2$ | Cu | 4.6 | ~5 | 2.6 | ≈70 |
| CsFe$_2$As$_2$ | Cu | 6.6 | ~4 | 1.8 | ≈30 |
| | | 8 | ~4.5 | | ≈50 |
| | | 5.6 | ~7 | | ≈40 |
| | Ag-soft | 7 | ~4 | | ≈20 |